\documentclass[a4paper,11pt]{article}
\usepackage{pos}
\usepackage{units} 
\usepackage{cleveref} 
\usepackage{subfigure,wrapfig} 
\title{Searches for ultra-high energy photons and neutrinos with the Pierre Auger Observatory}
\ShortTitle{UHE Photons and Neutrinos with Auger}

\manuallySeparateAuthors
\author*[a]{Nicolás~Martín~González}
\author[b]{ on behalf of the Pierre Auger}
\author{ Collaboration}

\affiliation[a]{Inter-University Institute for High Energies, Université Libre de Bruxelles, Belgium}
\affiliation[b]{Observatorio Pierre Auger, Av. San Martín Norte 304, 5613 Malargüe, Argentina}
\affiliation[]{Full author list: \url{http://www.auger.org/archive/authors_2022_07.html}}
\emailAdd{spokespersons@auger.org}

\abstract{The Pierre Auger Observatory is the largest astroparticle experiment in operation. 
Complementary to the measurements of the charged ultra-high energy (UHE) cosmic rays, it provides a very good sensitivity to the detection of UHE photons and neutrinos.
Since the photon and neutrino fluxes are correlated to the acceleration mechanisms of charged particles, searches for these neutral particles enhance the multi-messenger understanding of UHE cosmic-ray sources and of transient astrophysical phenomena.
In addition, searches for diffuse fluxes may bring information about exotic scenarios such as the decay of hypothetical super-heavy dark matter in the Galactic halo.
In this contribution, we present an overview of the current UHE photon and neutrino searches at the Observatory and discuss the most recent results.
We report on stringent limits to the UHE photon and neutrino diffuse and point-like fluxes above \unit[$10^{17}$]{eV}, which lead to strong constraints on theoretical models describing the nature of dark matter candidates and the sources of the most energetic particles in the Universe.}

\FullConference{%
  *** 27th European Cosmic Ray Symposium - ECRS ***\\
  *** 25-29 July 2022 ***\\
  *** Nijmegen, the Netherlands ***
}

\begin{document}

\renewcommand{\printHeadAuthors}{Nicolás González}

\maketitle

\section{Introduction}
\label{sec:Intro}

The sources of the ultra-high energy (UHE) cosmic rays are still elusive. Direct information on the cosmic-ray acceleration mechanisms can be inferred from the observation of neutral messengers, such as photons and neutrinos, generated in the interactions of cosmic rays with surrounding matter in their sources. Additionally, diffuse cosmogenic fluxes of photons and neutrinos are produced during the propagation of cosmic rays in the interaction with microwave low-energy photons. Since these fluxes depend on the composition and maximum energy of cosmic rays at the sources, their observation implies constraints on the nature of the cosmic-ray sources.

Photons and neutrinos are complementary messengers. On one hand, UHE photons have a limited propagation legth due to the interaction with background radiation fields. Thus, sources farther than a few Mpc are obscured, making UHE photons excellent probes of the local Universe~\cite{LHAASO2021}. For instance, an observation of photons clustered in the direction of the Galactic center could highlight dark matter particles decays~\cite{Alcantara2019}. On the other hand, neutrinos travel unperturbed along cosmological distances. They represent a powerful tool to probe the cosmological evolution and energy distribution of extragalactic objects~\citep{Collaboration2019b}.

The flux of UHE cosmic rays is too low to be directly measured with space instruments. Nevertheless UHE particles interact with Earth's atmosphere initiating extensive air-showers that are measured with large ground-based detectors. Thanks to its instrumented area of \unit[$3000$]{$\text{km}^2$}, the Pierre Auger Observatory~\cite{Collaboration2015f} has an outstanding exposure to UHE photons and neutrinos. Its hybrid detection principle encompasses the measurements of the longitudinal shower development with a Fluorescence Detector (FD), operated on clear and moonless nights, and the measurements of the lateral shower profile at ground level by the Surface Detector (SD), with a duty cycle of almost $100\%$. The SD consists of $\sim1600$ water-Cherenkov detectors arranged on a \unit[$3000$]{km$^2$} triangular grid, separated by \unit[$1500$]{m}, and overlooked by $24$ FD telescopes. A nested SD grid of $60$ stations separated by \unit[$750$]{m} and three high-elevation FD telescopes overlooking them are employed to measure cosmic rays from below the second knee up to the ankle of the cosmic-ray energy spectrum~\cite{Collaboration2021g}.
	
In this contribution, we present a summary of the recent UHE photon and neutrino searches in the Pierre Auger Observatory. Integral upper limits on the diffuse fluxes above \unit[$\sim10^{17}$]{eV} are reported. The sensitivity to steady and transient point-like neutrinos sources is discussed in the light of the multi-wavelength observation of astrophysical events.

\section{Search for UHE photons}
\label{sec:Photons}

Air-showers initiated by photon primaries have an almost pure electromagnetic (e.m.) development. The smaller multiplicity of the e.m. interactions compared to the hadronic ones causes the shower to develop deeper in the atmosphere. Consequently, the atmospheric depth of the maximum development, $X_\text{max}$, is expected to be larger for a photon-initiated shower compared to a hadronic one. Photon-initiated showers also contain fewer secondary muons which causes a steeper lateral fall-off and a slower rise of the signal registered by ground detectors than for the case of hadronic-initiated showers~\cite{Risse2007}.


The complementary detection techniques of Auger can be employed to measure the distinctive features of photon-initiated showers. Three discrimination methods, optimized in different energy ranges, are tailored and applied to the recorded data. The measured observables are combined into a single discriminator using a multi-variate method. The selection of candidates is based on the median of this distribution, leading to a $50\%$ signal efficiency. The number of candidates is computed and compared to the expected background contamination in order to assess their significance.

\begin{figure}
\subfigure[\label{fig:lowE_ph_BDT}]{\includegraphics[width=0.33\textwidth]{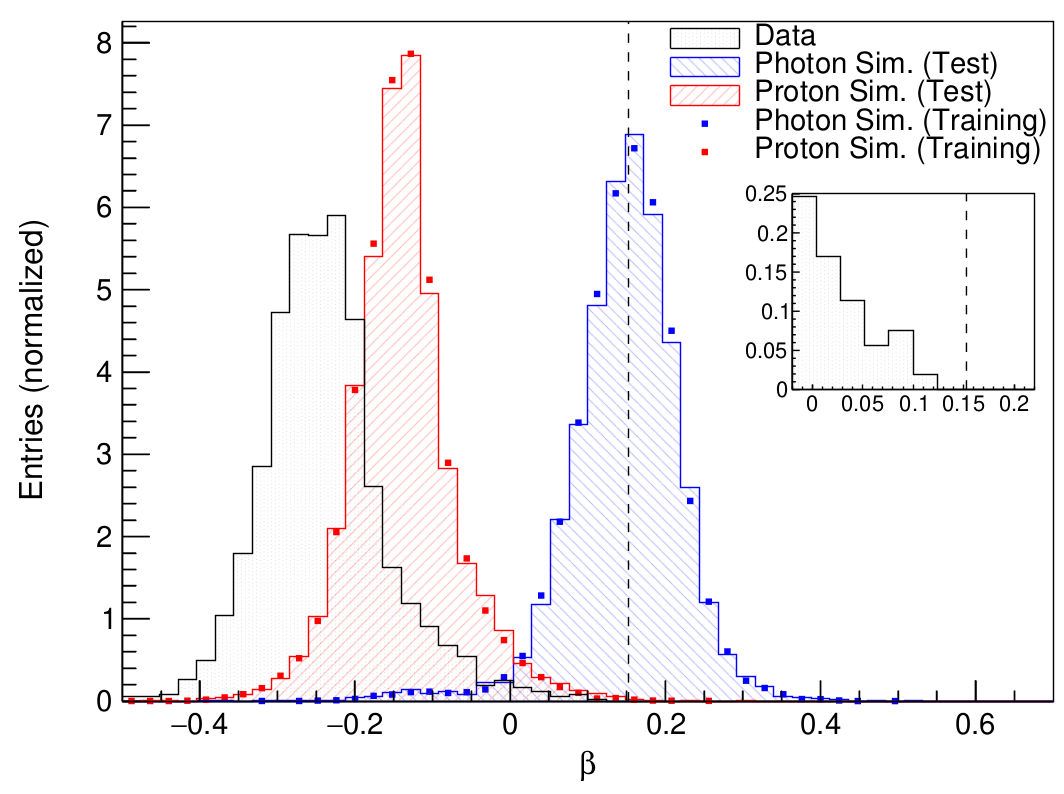}}
\subfigure[\label{fig:midE_ph_Fisher}]{\includegraphics[width=0.33\textwidth]{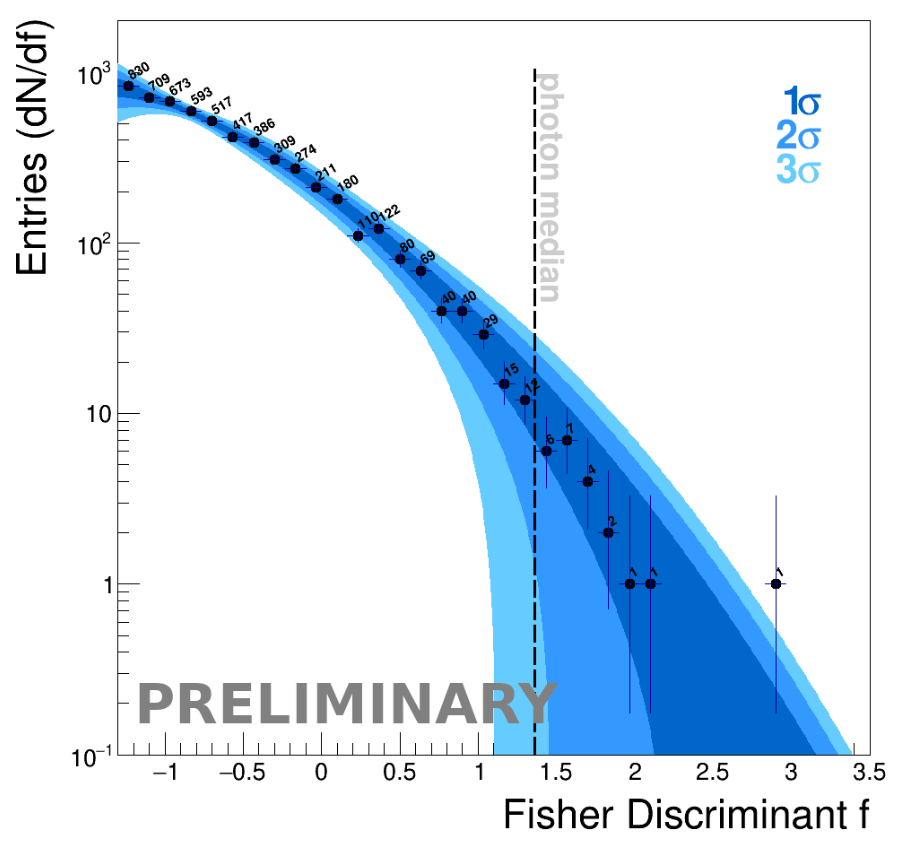}}
\subfigure[\label{fig:highE_ph_Principal}]{\includegraphics[width=0.33\textwidth]{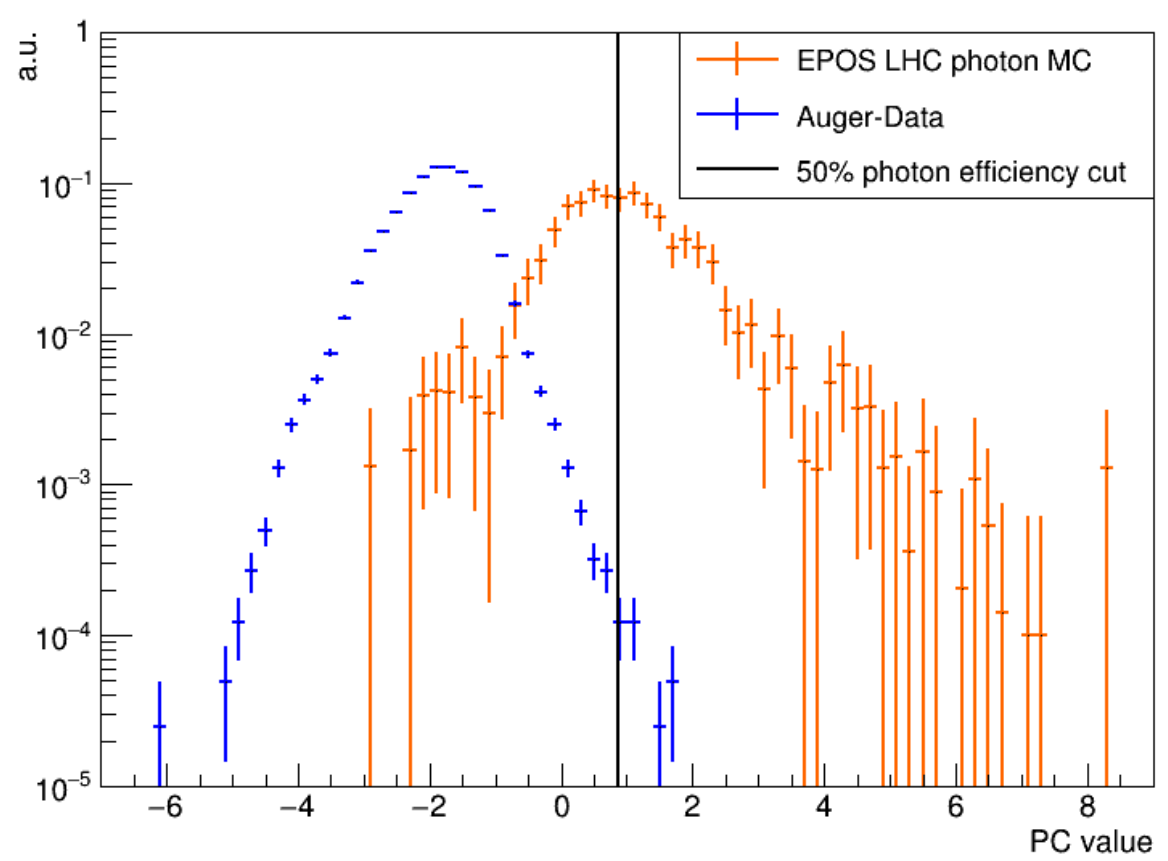}}
\caption{(a) The distributions of the BDT discriminator $\beta$ combining the observables $X_\text{max}$, $S_b$ and $N_\text{trig}$ above \unit[$2\times10^{17}$]{eV}. The photon (proton) samples are shown in blue (red), while data is shown in black. The inset shows the data distribution near the photon candidate cut (dashed line). (b) The tail of the Fisher distribution combining $X_\text{max}$ and $F_\mu$ for events above \unit[$10^{18}$]{eV} (black bullets) near the candidate cut (dashed line). The blue-shaded region represents the expected background contamination, derived from $5\%$ of the data. (c) The normalized distributions of the PCA estimator combining $L_\text{LDF}$ and $\Delta$ above \unit[$10^{19}$]{eV} for simulated photons (orange) and the search data set (blue). The black line represents the candidate cut.}
\end{figure}

	
The hybrid data acquired by the low-energy extensions of the SD and FD are explored to discriminate a photon signal among the hadronic background above \unit[$2\times10^{17}$]{eV}~\cite{Collaboration2022b}. For this task, the direct measurements of $X_\text{max}$ are complemented by the observable $S_{b=4}$ defined as $S_b = \sum_i S_i\times\left(\frac{r_i}{\unit[1000]{\text{m}}}\right)^b$ that combines the measured signals $S_i$ recorded by stations located at a distance $r_i$ from the shower axis. The number of triggered stations per event, $N_\text{trig}$, $X_\text{max}$ and $S_b$ are combined via a Boosted Decision Tree (BDT) method trained using simulated proton events as background. The distribution of the output estimator $\beta$ is displayed in \cref{fig:lowE_ph_BDT}. The background contamination at the candidate cut is $(0.09\pm0.03)\%$, which translates to an expectation of $1.98\pm0.66$ background events in the final data set. However, no events have been tagged as candidates.
	

Above \unit[$10^{18}$]{eV}, the photon search is driven by the measurements of $X_\text{max}$ combined with an observable sensitive to the muon content of the air-showers, $F_\mu$~\cite{Savina2021}. The latter is estimated with a method based on the Universality of air-showers, which relies on the idea that the energy spectra of the secondary particles in a shower depend mostly on the primary energy, the stage of shower development and the primary mass~\cite{Lipari2009}. The predicted signal on the SD stations $S_\text{pred}$ is modeled as a superposition of the muonic components and three e.m. components, classifying the particles depending on the progenitor species~\cite{Ave2017}. The relative contributions of each component depends on the mass-sensitive $F_\mu$ parameter. Thus by matching $S_\text{pred}$ to the recorded signal the parameter $F_\mu$ is estimated. A Fisher discriminant is built from $X_\text{max}$ and $F_\mu$. The expected background contamination is derived with a sub-set of $5\%$ of the data and scaling the normalization to the size of the full data sample. The data distribution is shown in \cref{fig:midE_ph_Fisher}. Twenty-two events are marked as candidates, however they are consistent with the expected number of $30\pm15$ background events.
	
		
Above \unit[$10^{19}$]{eV}, the SD events are scrutinized employing two composition-sensitive estimators~\cite{Collaboration2022a}. The first one, $L_\text{LDF}$, is based on the deviation of the measured SD signals from the average data-driven lateral distribution function (LDF). The second observable, $\Delta$, characterizes the deviation of the signal risetime, defined as the time difference between the $10\%$ and the $50\%$ quantiles of the SD signal traces, from the average expectation in data. Photon-induced showers are expected to develop steeper LDFs and signals with a slower rise. The two observables are combined with a Principal Components Analysis (PCA) into a single discriminator, whose distributions for simulated photon and data events are displayed in \cref{fig:highE_ph_Principal}. Although $16$ events pass the candidate cut, further analyses using dedicated proton simulations are not significantly excluding the background hypothesis.
	


\begin{wrapfigure}{R}{0.7\textwidth}
\begin{center}
\includegraphics[width=0.7\textwidth]{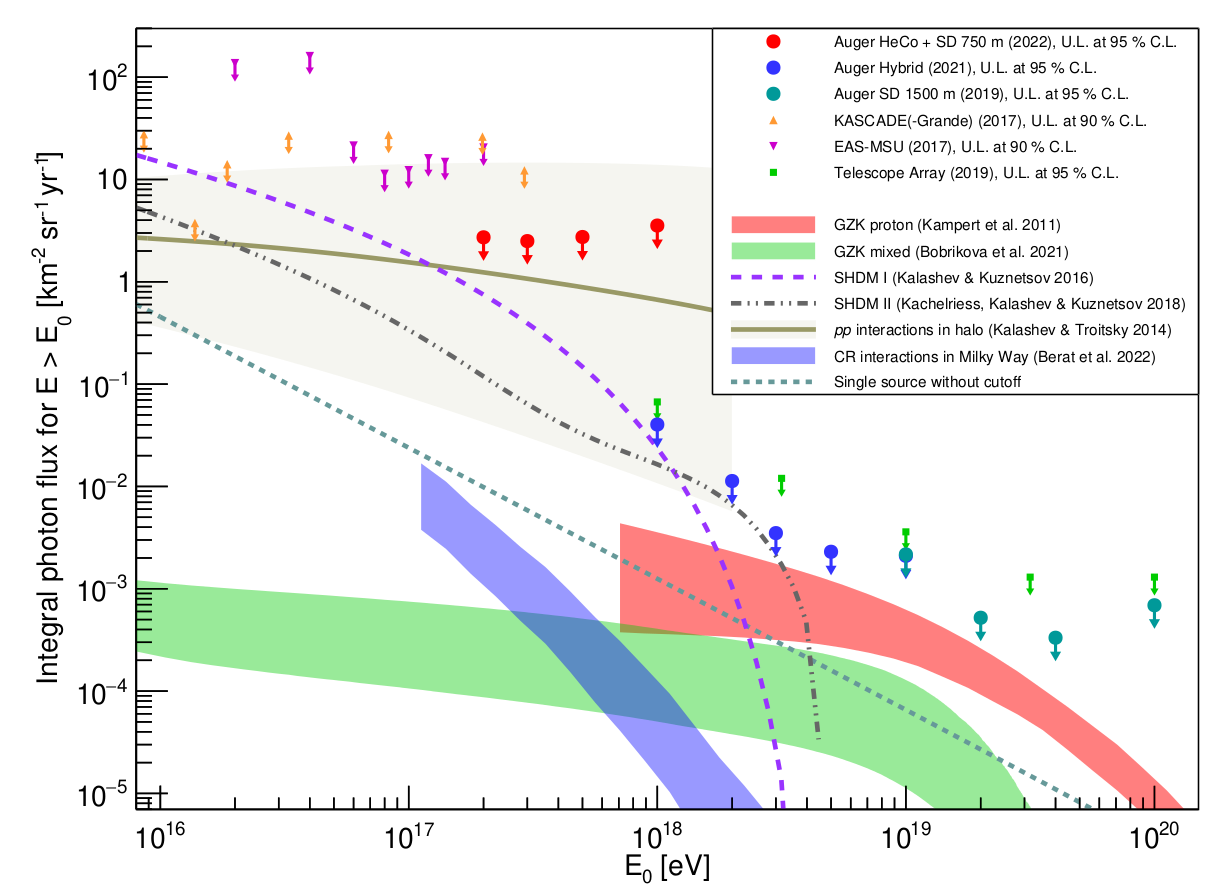}
\end{center}
\caption{\label{fig:ph_UL} The upper limits on the integral UHE photon flux established by Auger (circles)~\cite{Collaboration2022b}. Shown also are previous results by various experiments in triangles and squares: KASCADE-Grande~\cite{KASCADEGrande2017}, EAS-MSU~\cite{Fomin2017} and Telescope Array~\cite{Abbasi2019}. Filled bands denote the range of the expected cosmogenic photon fluxes assuming different cosmic-ray composition. The predicted fluxes from the decay of dark-matter particles and from the interaction between cosmic-ray primaries and Galactic matter are shown as dashed and solid lines, respectively.}
\end{wrapfigure}

Since these results are consistent with the expectations from the hadronic background, upper limits have been placed on the integral photon flux above a minimum energy $E_0$. They are determined considering the number of candidate events at a confidence level of $95\%$~\cite{Feldman1998} and the integrated exposure to photon primaries computed from simulations. As shown in \cref{fig:ph_UL}, the upper limits derived with Auger data are the most stringent ones across three decades in energy.

Diffuse photon fluxes are guaranteed as the result from the interaction of UHE cosmic rays with background radiation fields~\cite{Bobrikova2021}, shown as red and green shaded bands in \cref{fig:ph_UL}. An improvement of one order of magnitude in the experimental sensitivity is required to reach the expectation assuming a 	primary composition compatible with the latest Auger results~\cite{Collaboration2017f}. Another cosmogenic flux can be produced in the interactions of UHE cosmic rays with the interstellar Galactic matter~\cite{Berat2022}, which is larger than the aforementioned component below \unit[$10^{18}$]{eV} (blue band in \cref{fig:ph_UL}). Predictions from various phenomena may emerge above the cosmogenic fluxes below about \unit[$10^{18}$]{eV}. For example, the interaction between UHE cosmic rays and hot gas in the Galactic halo can produce pions that decay into neutrinos and photons. By requiring that the observed astrophysical neutrino flux~\cite{IceCube2020} is entirely generated by this mechanism, the photon counterpart can be reproduced (olive green in \cref{fig:ph_UL}). The reported upper limits are constraining these last assumptions. Lastly, as the sensitivity of current photon searches increases, it will be possible to also reach the predictions from the decay of super-heavy dark matter~\cite{Anchordoqui2021} (dashed grey and purple lines), thus constraining the phase space of mass, lifetime and decay channels of these putative particles.

		
\section{Search for UHE neutrinos}
\label{sec:Neutrinos}

Neutrinos can initiate air-showers after traversing a large amount of matter. This condition is met for a very-inclined incidence with respect to the vertical, either through the Earth's crust or through the atmosphere~\cite{Auger2011}, giving birth to two neutrino detection channels. In the earth-skimming (ES) channel, $\tau$ neutrinos are expected to be observed through the detection of showers induced by the decay of emerging $\tau$ leptons which are created by a neutrino interaction in the Earth. In the ES case, the primary zenith angle is constrained between $\theta=90^\circ$ and $95^\circ$ for which the showers initiated by $\tau$ leptons efficiently trigger the SD. In the second case, down-going (DG) neutrinos of all flavors interacting in the atmosphere can induce an air-shower close to the ground when arriving with a zenith angle $\theta>60^\circ$.


The main challenge in detecting UHE neutrinos is to identify a neutrino-induced shower in the hadronic background. Inclined air-showers reaching the ground traverse enough matter to have the e.m. component absorbed. Thus just the muonic component reaches the detectors. On the other hand, neutrino-induced showers are initiated at a much lower altitude and arrive to the ground with a prominent e.m. component spread over hundreds of nanoseconds. The discrimination method is designed to ensure that the inclined showers have a large e.m. component at ground level compared to the background expectation. It is based on the Area-over-Peak (AoP), defined as the ratio of the integral of the SD trace to its peak value. In background horizontal showers, the SD signal traces are composed of faster and shorter pulses, so with a smaller AoP, than for neutrino-induced showers.

\begin{figure}[!tb]
\subfigure[\label{fig:AoP_ES}]{\includegraphics[width=0.495\textwidth]{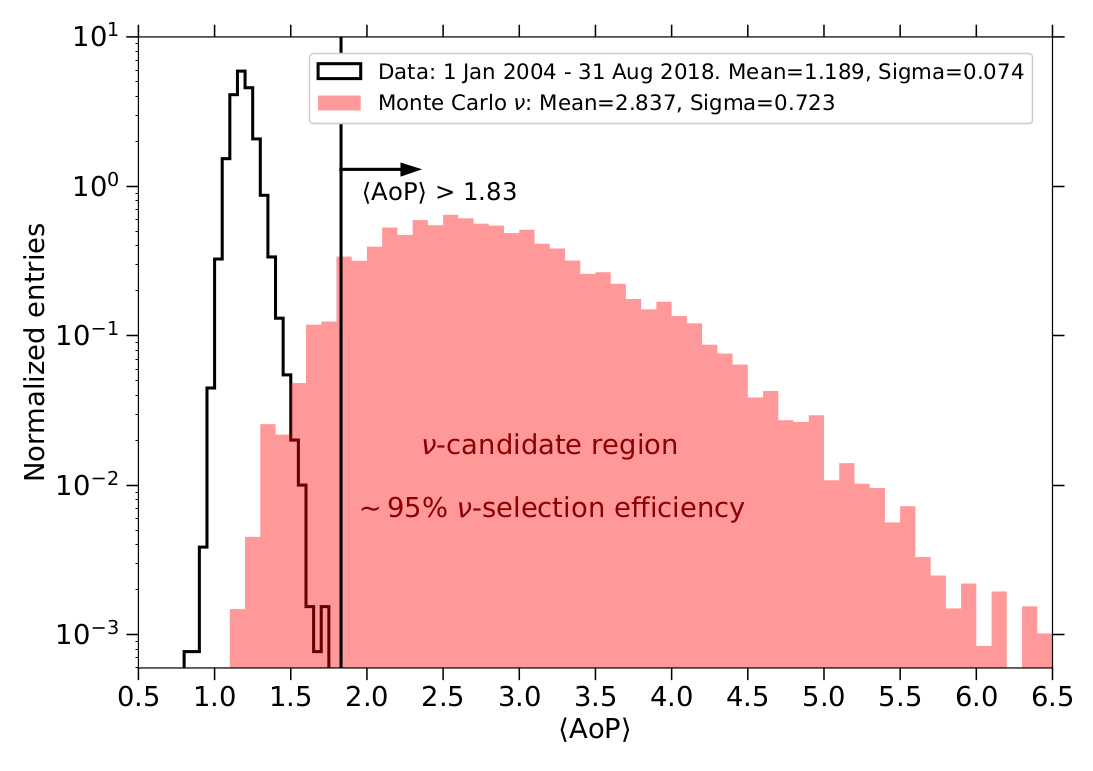}}
\subfigure[\label{fig:nu_exposure}]{\includegraphics[width=0.495\textwidth]{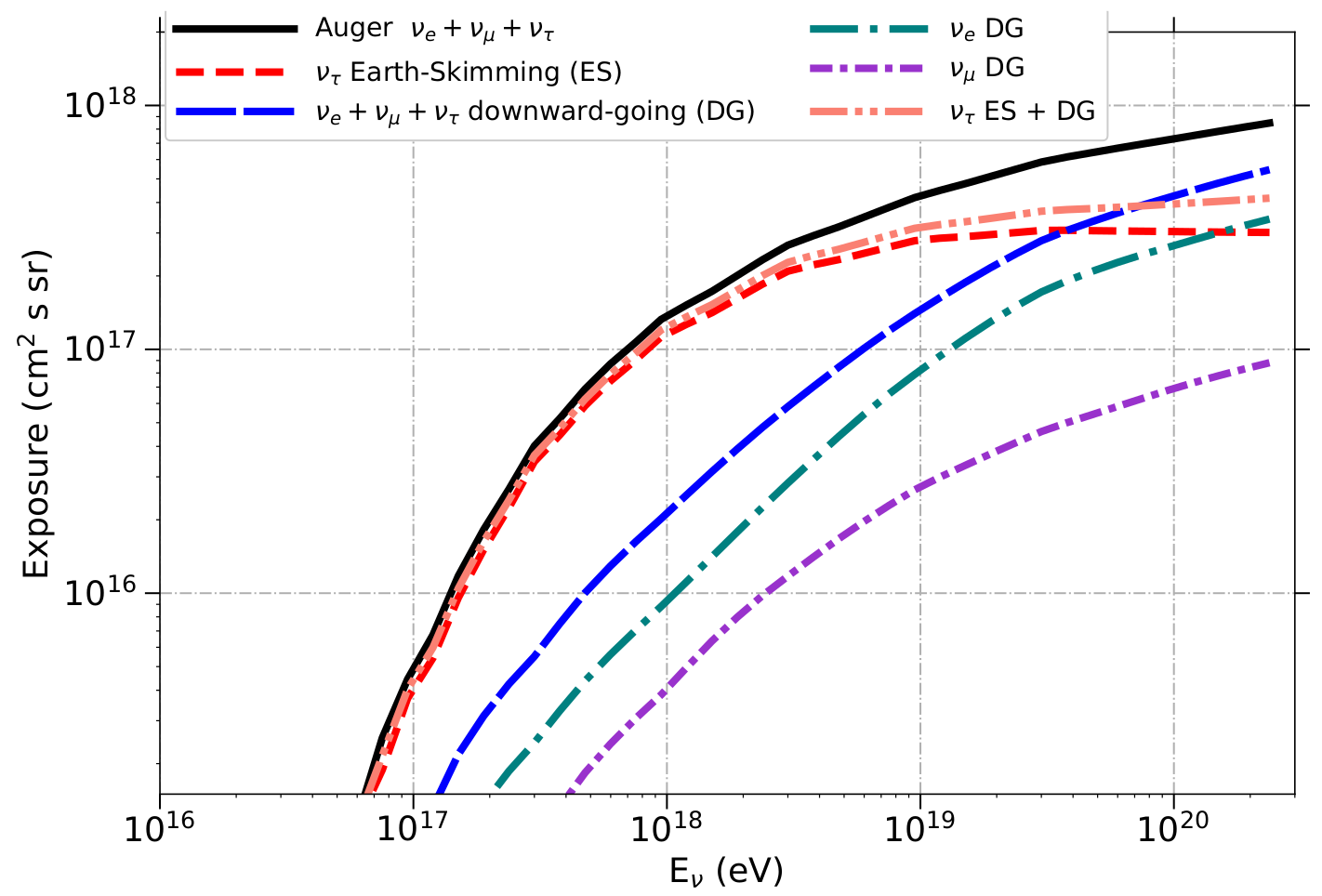}}
\caption{(a) The distribution of $\langle\text{AoP}\rangle$ for the search data-set (black) and simulated ES neutrino events (red). (b) The SD neutrino exposure as a function of the primary energy for the two detection channels~\cite{Collaboration2019b}.}
\end{figure}

The distribution of the average AoP in each event is shown in \cref{fig:AoP_ES} for real data and simulated showers induced by ES $\tau$ neutrinos. The cut on $\langle\text{AoP}\rangle$ is optmized with a data sub-set as to ensure a $95\%$ $\nu$-selection efficiency with a background contamination such that less than one false-positive event is expected in $50$ years of data~\cite{Collaboration2019b}. This implies that the ES neutrino sensitivity is only limited by the observation time and is not affected by the background contamination. For DG showers, a Fisher discriminant combining several observables constructed from the AoP values of individual stations is used. The candidate cut is set on the Fisher discriminant requiring a background contamination of less than one event in $50$ ($20$) years for events with zenith angles $75^\circ < \theta < 90^\circ$ ($60^\circ < \theta < 75^\circ$). The $\nu$-selection efficiency is estimated as $85\%$ and $72\%$, respectively~\cite{Collaboration2019b}, due to the less attenuated e.m. component at ground in hadronic showers with decreasing zenith angle.

The SD exposure to a neutrino flux is calculated with simulations, convoluting the energy-dependent detection efficiency and the time evolution of the acceptance. In \cref{fig:nu_exposure}, we show the exposure contribution for each neutrino flavor and detection channels. The $\nu_\tau$ flavor contributes the most to the total exposure, mainly because of the enhanced sensitivity of the ES channel. At the highest energies, the emerging $\tau$ lepton decays at a high altitude such that the up-going shower may not trigger the SD, softening the exposure of the ES channel. In the case of the DG channel, a larger contribution to the exposure comes from $\nu_e$ and $\nu_\tau$ due to the larger fraction of neutrino energy transferred to the e.m. component of the induced shower compared to the case of $\nu_\mu$.

\begin{figure}[!tb]
\subfigure[\label{fig:nu_UL}]{\includegraphics[width=0.495\textwidth]{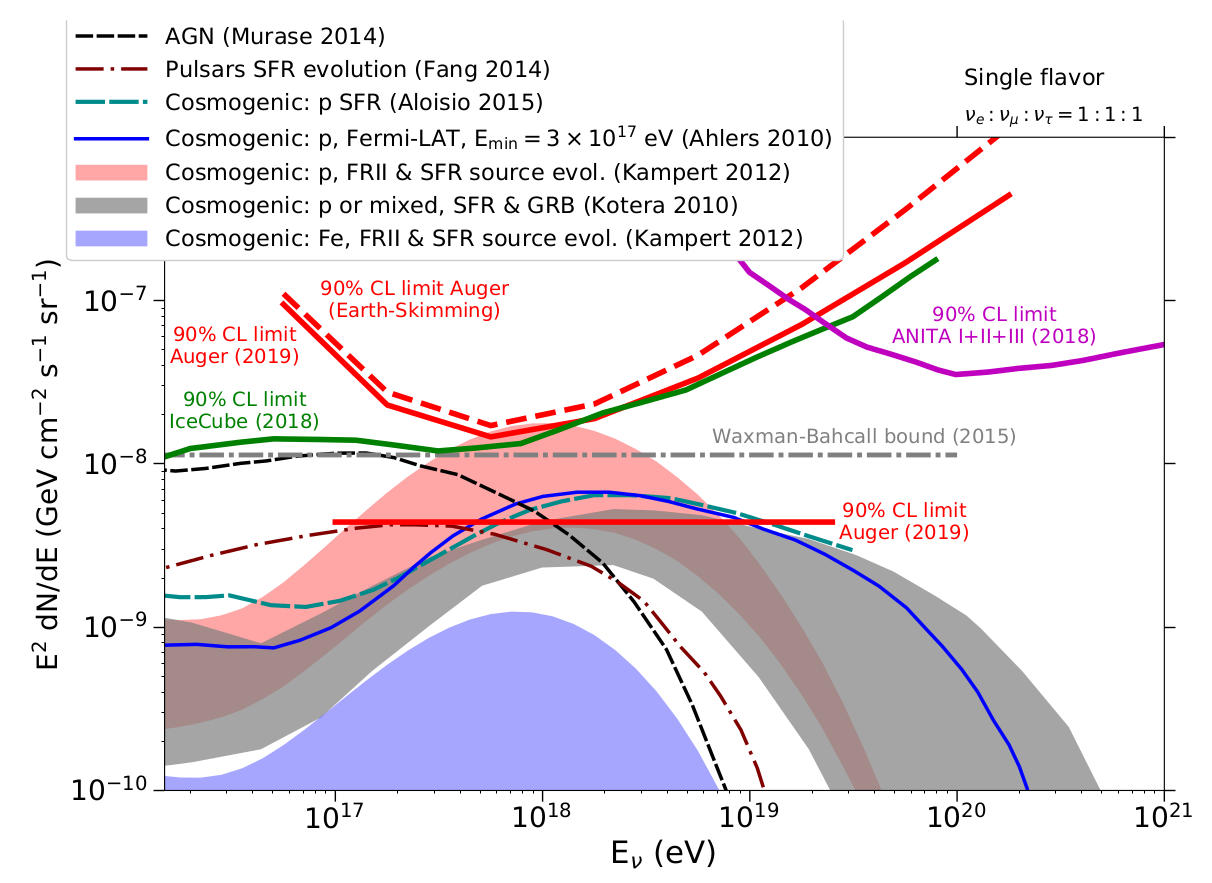}}
\subfigure[\label{fig:nu_UL_declination}]{\includegraphics[width=0.495\textwidth]{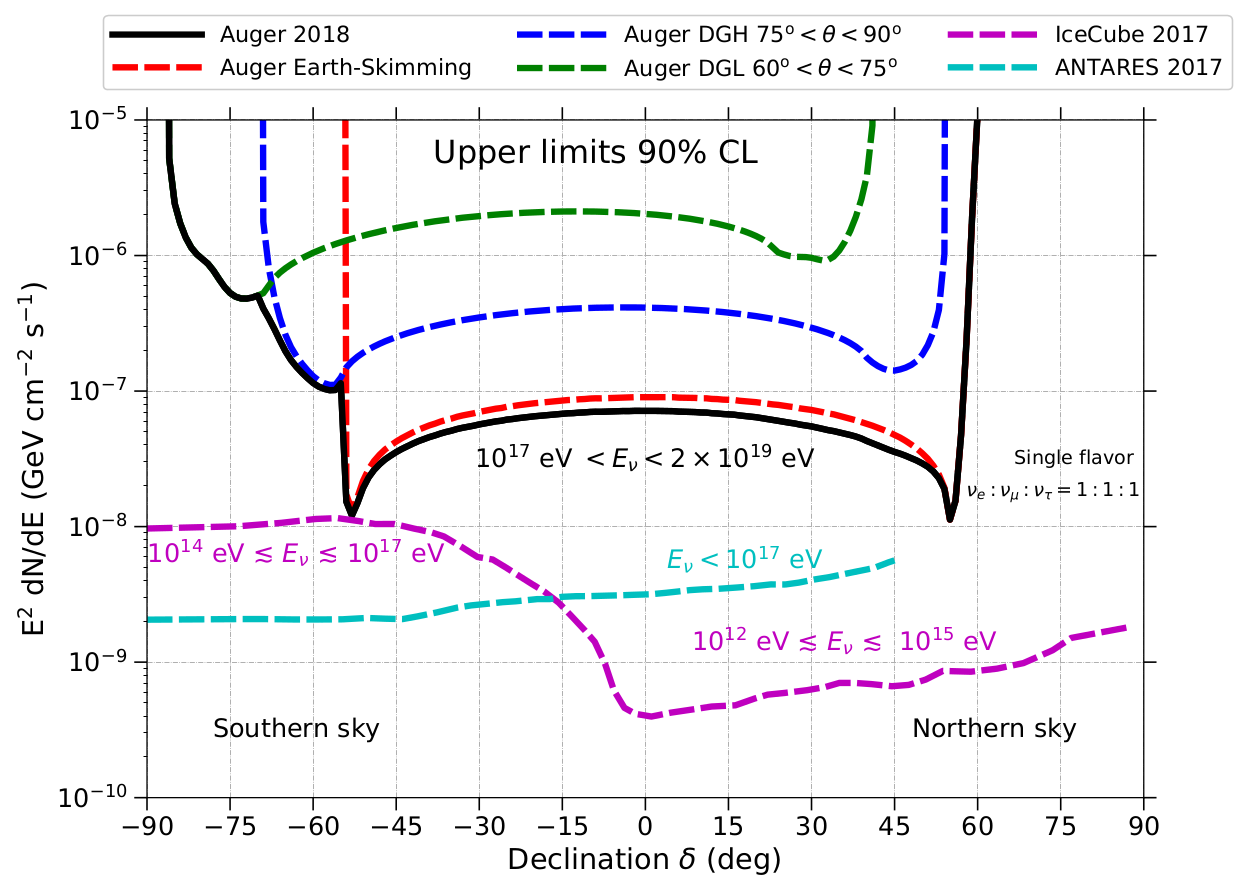}}
\caption{The upper limits at $90\%$ confidence level on the diffuse UHE neutrino flux obtained with Auger SD. (a) Diffuse differential and integral limits (red lines) are shown accompanied by those estimated with data by other observatories and predicted by several cosmogenic and astrophysical models~\cite{Collaboration2019b}. All limits and fluxes are converted to single flavor. (b) Upper limits as a function of the source declination $\delta$. Also shown are the limits for IceCube and ANTARES. Note the different energy ranges where the limits of each observatory apply. Figures extracted from~\cite{Collaboration2019d}.}
\end{figure}

No neutrino candidates have been observed in the acquired data, thus upper limits to the diffuse neutrino flux are established. As displayed in \cref{fig:nu_UL}, the best sensitivity is achieved around \unit[$10^{18}$]{eV}, beyond the energies accessible with IceCube. The energy-integrated upper limit is $4.4\times10^{-9}\,$GeV$\,$cm$^{-2}\,$s$^{-1}\,$sr$^{-1}$. It mostly applies between \unit[$10^{17}$]{eV} and \unit[$2.5\times10^{19}$]{eV} where $90\%$ of the total event rate is expected. Given this upper limit, models assuming sources that accelerate only protons with a strong redshift evolution are strongly disfavored, while a three(six)-fold increase in the exposure would be needed to assess models assuming a mixed (pure iron) composition.

After the birth of gravitational wave astronomy~\cite{LIGO2017}, there is much interest in detecting UHE neutrinos in transient phenomena to constrain possible production mechanisms. Given the angular requirements of the ES and DG channels and the periodically-changing field of view, the neutrino identification efficiency of Auger is dependent on the direction in the sky where the search is performed. In \cref{fig:nu_UL_declination}, the upper limits as a function of a steady point-source declination $\delta$ are presented~\cite{Collaboration2019d}. Auger is sensitive to sources located at a broad declination range from $-85^\circ$ to  $60^\circ$, being the most sensitive experiment above \unit[$10^{18}$]{eV} for sources in the Northern hemisphere. This region in the sky cannot be searched for at these energies by IceCube because of the opacity of the Earth to neutrinos in those directions when seen from the South Pole. The ES channel provides the best sensitivity for two specific declinations corresponding to the longest transit of the source in its field of view.

In addition to the declination dependence, the sensitivity to transient sources is crucially dependent on the ocurrence time interval of such event. After a detection of a relevant astrophysical event, a follow-up campaign of observations across the e.m. spectrum is triggered. For instance, the gravitational wave (GW) event GW170817 caused by a neutron star merger~\cite{LIGO2017b} was transiting the ES field-of-view at the detection time, so that the most stringent upper limits at UHE were provided by Auger in a short time window around the event timestamp. The upper bounds to the neutrino (and photon) fluxes were consistent with the expectations from an off-axis short gamma-ray burst~\cite{ANTARESCollaboration2017}, demonstrating the science potential of follow-up observations. Besides automated searches upon multi-messenger alerts, the Auger collaboration has conducted stacking analyses of photon and neutrino emission following GW events~\cite{Schimp2021,Ruehl2021}.

\section{Conclusions}
\label{sec:Conclusions}

Thanks to its large exposure and hybrid base design, the Pierre Auger Observatory can be used to achieve an unrivalled exposure to UHE photons across three decades in energy above \unit[$2\times10^{18}$]{eV} and a background-free sensitivity to diffuse neutrino fluxes, only limited by the observation time. Limits on the fluxes of neutral particles are used to constrain from the nature of cosmic-ray sources and their cosmological evolution to the lifetime and mass of hypothetical dark matter particles in the Galactic halo. In addition, the surface array of Auger has an excellent sensitivity to steady point UHE neutrino sources with nearly a full sky coverage and to transient events, specially if they are located in the ES field-of-view. Therefore, Auger is in an excellent position to contribute to the new era of multimessenger astronomy by looking for UHE neutrinos in correlation with the detection of gamma rays or GWs.

The bounds on UHE photons and neutrinos will become stronger with more and enhanced statistics acquired with AugerPrime, the upgrade of the Pierre Auger Observatory~\cite{Castellina2019}. A major part of this upgrade is the installation of scintillators detectors on top and buried beneath each SD station, aiming at improving the separation between the e.m. and muonic shower components, and antennas to measure the radio emission of air-showers. Naturally, this will lead to a better separation between showers initiated by neutral particles and those initiated by charged cosmic rays.

\renewcommand*{\bibfont}{\small}


\end{document}